\documentclass[amsmath,amssymb,nobibnotes]{revtex4-2}
\usepackage[T1]{fontenc}
\usepackage[utf8x]{inputenc}
\usepackage{latexsym}
\usepackage{graphicx}
\usepackage{amsfonts,amsbsy,amssymb,amsthm}
\usepackage{amsopn}
\usepackage{color}
\usepackage{ulem}

\begin {document}
\title{Information scrambling and chaos induced by a Hermitian Matrix}

\author{Sven Gnutzmann\(^1\) and Uzy Smilansky\(^2\)}
\affiliation{\(^1\)School of Mathematical Sciences, University of
Nottingham, United Kingdom}
\affiliation{\(^2\)Department of Physics of Complex Systems,
The Weizmann Institute, Rehovot, Israel }

\begin{abstract}
Given an
arbitrary \(V \times V\) Hermitian matrix,
considered as a finite discrete quantum Hamiltonian,
we use methods from graph and
ergodic theories to
construct a \textit{quantum Poincar\'e map} at energy \(E\) and a
corresponding stochastic
\textit{classical Poincar\'e-Markov map} at the same energy
on an appropriate
discrete
\textit{phase space}.
This phase space
consists of the directed edges of a
graph with \(V\) vertices that are in one-to-one correspondence with the
non-vanishing off-diagonal elements of \(H\).
The correspondence between quantum Poincar\'e map and classical
Poincar\'e-Markov map is
an alternative to the standard quantum-classical
correspondence based on a classical limit \(\hbar \to 0\). Most
importantly it can be constructed where no such limit exists.
Using standard methods from ergodic theory
we then proceed to define
an expression for the
\textit{Lyapunov  exponent} \(\Lambda(E)\) of the classical map.
It measures the rate of loss of classical information in the dynamics
and relates it to the separation of stochastic \textit{classical
trajectories} in the phase space. We suggest that
 loss of information in the underlying
classical dynamics is an indicator for quantum information scrambling.
\end{abstract}


\maketitle

\section{Introduction}

Quantum information scrambling and operator growth is currently
investigated in many fields such as black holes theory \cite{hayden,sekino,Shenker,Maldacena},
 condensed matter many-body physics
 \cite{Kitaev}, and quantum information theory \cite{qit}.
 They brought together developments  in quantum information and
quantum chaos, and sparked a large overarching interest that also lead to
experimental observations \cite{landsman,mi,harris}.
Mathematically, quantum  information scrambling is related to
seminal results such as the Lieb-Robinson (LR) quantum speed
limits \cite{lieb,chen}.
One of the main tools used  to quantify information scrambling
is the
so-called out-of-time-ordered commutators [OTOC],
initially introduced  in condensed matter physics \cite{Larkin}
(see \cite{scholarpedia}
for an overview and further references). The exponential
growth of OTOC  is assumed to signal
operator growth and information scrambling.

In classical  mechanics, information scrambling  is associated with
classical chaos.
In contrast with its quantum counterpart, it is a well defined concept
where both
physical and mathematical tools and methods are  available to study and
quantify chaotic loss of information \cite{ChaosBook,Ott}.
They provide a solid explanation why deterministic mechanical systems
might not be predictable which justifies their study using statistical
rather than deterministic methods.

No wonder therefore that the quantum correspondence principle was tried
to link the study of quantum chaotic features with an underlying classical concepts.  Indeed, in \cite{Shenker,Maldacena}
the exponential growth rate of OTOC was suggested  as the
`quantum butterfly effect' and
`quantum Lyapunov exponent'.
However, this notion is rather limited since exponential growth of
OTOCs has been found even in classically
regular systems \cite{otoc-counter1,otoc-counter2}.
The study of quantum systems which are  defined by quantizing
classical chaotic  systems was the subject of the field of quantum chaos
\cite{ChaosBook,Gutzwiller,QSoC}.
The main tool was the study of the quantum systems in the
limit $\hbar \rightarrow 0$. Clearly, not all the classical quantities
could be transplanted to the quantum theory.
Yet,  one could still ask
what are the finger-prints of classical chaos in the
quantum description.
The BGS conjecture \cite{BGS} e.g.,
suggests that the spectral statistics of `typical' classically
chaotic quantum systems follow the statistics provided by
Random Matrix Theory (RMT).
This conjecture does not apply in the
other direction -- namely, a quantum system may display RMT
statistics without
having a chaotic classical counterpart.
As an example consider two different
random matrix ensembles:
The Wigner-Dyson ensembles where all the matrix elements are randomly
distributed, and the Dumitriu-Edelman ensembles (G\(\beta\)E)
\cite{Edelman} with random entries forming tridiagonal matrices. 
They share precisely the same spectral distribution functions. 
However, while eigenvectors of the former
are uniformly distributed (an  indicator
for chaos via the Shnirelman theorem \cite{Shnir}), those
of the latter display a transition from localised eigenfunctions
(implying suppressed chaos) for \(\beta \le 2\)
values to extended eigenfunctions otherwise \cite{BFS}. 
We will return to this example in the sequel. 
In general, a single quantum signature of chaos -- such as
spectral statistics --  as indicator might lead to
an incomplete or insufficient picture of the chaoticity of the quantum
system.

The purpose of this note is to introduce a new method for
quantifying the degree of information loss (or  chaoticity) induced by
quantum evolution. Being the first time this approach is presented, its
ideas and main results will be explained, leaving applications and further
results to later publication.

The system to be discussed is driven by a quantum Hamiltonian
represented as a finite
matrix. The quantum evolution reduces in the semi-classical limit to a
discrete stochastic classical dynamics expressed in terms
of a Poincar\'e-Markov map.
Ergodic theory is then used to define a classical Lyapunov exponent.
It describes the flow and loss of information in the classical dynamics
and relates it to the
deviation of (stochastic) trajectories in a discrete phase space which
is constructed specifically for the Hermitian matrix of interest. 

The new approach builds upon ideas, results and methods from
spectral graph theory and ergodic theory. It uses the results of a
recent work which provides an energy dependent
unitary map \(U(E)\) -- to which we refer as  the
\textit{Quantum Poincar\'e map} --
to a Hermitian matrix \(H\) in a non-trivial
way such that an underlying graph structure is
obeyed \cite{US2007,GnuSmi}.
This Quantum Poincar\'e map acts on an
associated phase space -- and the classical dynamics is obtained
by replacing quantum transition amplitudes by their absolute squares.
We call this the Poincar\'e-Markov map
as it is a stochastic matrix that generates a Markov process.
A similar
quantum-classical correspondence was
introduced \cite{KS-prl,KS-anp} for quantum graphs,
(see \cite{QG-review} and references therein). It was used e.g.,
to set criteria to determine  which  graphs would display
RMT spectral statistics
\cite{KS-prl,KS-anp, QG-review,  Gaspard, berkolaiko, tanner, pakonski, GnuAlt}.

A few examples show how
the classical Lyapunov exponent reflects localization properties
of wave functions.
Further applications and results as well as future research directions,
together with a discussion of  the relation to OTOC and the LR
speed limit \cite{lieb,chen} are deferred to the last section.

\section{The  classical dynamics associated with a matrix and
  the Lyapunov exponent}

The quantum system under study is governed
by a Hamiltonian which is represented as
a \(V\times V \) Hermitian matrix \(H_{v w}\)  with \(D\)
non-vanishing off-diagonal elements. 
Without loss of generality it is assumed that \(H\) is not
block-diagonal. 
The energy spectrum and eigenvectors satisfy,
\begin{equation}
  \sum_{w=1}^V H_{v w}\phi_w = E \phi_v
  \label{schrodinger_equation}
\end{equation}

The theory is  presented below in the following steps:
\begin{itemize}
\item[\textbf{Step 1}]
  Eq.~\eqref{schrodinger_equation}
  is rewritten  as a discrete
  Schr\"odinger equation with ``kinetic energy'' and ``local potential''
  terms on a graph with  \(V\)
  vertices and \(D\) directed edges. The set of vertices is naturally
  defined as the \textit{configuration-space} on the graph.
  The analogous classical dynamics is that of
  hopping between neighbouring vertices.
\item[\textbf{Step 2}]
  A discrete momentum on the graph is introduced so that
  the classical \textit{phase-space} is  the set
  of directed edges on the graph. The quantum evolution operator
  or \textit{Quantum Poincar\'e map}
  is expressed as a \(D\times D\)  unitary matrix \(U(E)\)
  in the directed-edge basis. The intimate connection
  between \(H\) and \(U(E)\) is evident since the
  spectrum of \(H\) consists of the zeros of
  \(\det(\mathbb{I}_D-U (E))\).
\item[\textbf{Step 3}]
  The corresponding classical
  dynamic -- the \textit{Poincar\'e-Markov} map --
  is expressed in terms of
  classical transition \textit{probabilities}.
  They are defined
as the absolute squares of the quantum
  transition \textit{amplitudes}
  which are the matrix elements of \(U(E)\).
  The matrix obtained this way is a bi-stochastic matrix
  which defines a Markovian evolution on the graph.
\item[\textbf{Step 4}]
  Finally, standard methods from ergodic theory
  are used to compute
  the Lyapunov exponent and its variance for the
  Poincar\'e-Markov map associated
  to \eqref{schrodinger_equation} at a given energy \(E\).
\end{itemize}

Many of the ideas and methods applied in this work were
discussed and used in other contexts. 
We harness them here in order to introduce the novel
approach to information scrambling and chaoticity which is to
be unfolded.

\vspace*{0.5cm}
\noindent
\textbf{Step 1: A discrete Schr\"odinger operator on an underlying graphs}\\
We
associate to the matrix \(H\) an underlying graph \(\mathcal{G}\)
with \(V\) vertices and adjacency matrix
\begin{equation}
	\hspace{-10mm}
  A_{vw} =
  \begin{cases}
  1 &   \text{if  \( H_{vw}\ne 0\) \ and \ \(v \neq w\),}\\
  0 & \text{else.}
  \end{cases}\
  \label{adjacency}
\end{equation}
The degree of a vertex \(v\) is denoted by \(d_v=\sum_{w=1}^V A_{vw}\).
The graph vertex set \(\mathcal V\) forms the
\textit{configuration space}.
The Hamiltonian \(H\)
can be written as a generalised tight-binding
Schr\"odinger operator \(H=-L+W\)  with a kinetic energy (Laplacian)
part \(-L\)
that describes the hopping
and a diagonal potential \(W\)
\begin{eqnarray}
  L_{vw}= \Gamma_v\delta_{vw}-H_{vw}(1-\delta_{vw})\ \    \text{and} \ \
  W_{vw}=\left(H_{vv}+\Gamma_v\right)\delta_{vw} . \nonumber
\end{eqnarray}
Here, \(\Gamma_v=\sum_{u\ne v}|H_{v u}|\) is known as \textit{Gershgorin parameter} \cite{Gershgorin}.
It will appear often in the sequel.
The Gershgorin circle theorem \cite{Gershgorin}  implies that \(-L\)
is a non-negative matrix.
If \(H=-A\) the Gershgorin parameters reduce to
\(\Gamma_v= d_v\) and
\(L\) to the standard combinatorial graph
Laplacian \cite{US2007,GnuSmi}. Classical trajectories in
configuration space are strings of connected vertices.

\vspace*{0.5cm}
\noindent
\textbf{Step 2: The phase space and definition of the
  unitary quantum map}\\
To define the corresponding
\textit{phase space},  recall that
in classical mechanics the momentum points from
the present point in configuration space
to its future position. 
\textit{A-priori}, any vertex \(u\) could  be the ``next'' vertex to the
starting vertex \(v\). 
However, the graph connectivity limits the
possible choices to the adjacent vertices where \(A_{uv}=1\). 
It is  natural to define the
\textit{momentum space} as
the vertex set  which can be reached from a given
vertex by a single hopping. Thus,  \(A\) defines the domain of allowed momenta.  
A directed pair of connected vertices forms a directed edge.
Thus, \textit{Phase space}
is the space of all directed edges. 
Their total number is \(D= \sum_{v,w=1}^V A_{vw}\). 
For a given directed edge \(e=(vw)\),  the \textit{origin} is \(w=o(e)\)
and the \textit{terminus} is \(v = \tau(e)\). 
Classical trajectories in phase space are strings of
connected directed edges \(e_i\) where \( o(e_{i+1})=\tau(e_i) \). 

The \textit{phase space} evolution
will now be expressed in terms of a unitary evolution operator
on  a \(D\) dimensional space of amplitudes
\(a_{vw}\) with \( (vw) \in \mathcal{D}\). 
On a given edge that connects \(v\) and \(w\)
the amplitudes \(a_{vw}\) and \(a_{wv}\) are
defined in terms of the vertex
amplitudes
\(\phi_v,\ \phi_w\)
of \eqref{schrodinger_equation}.
It is convenient to denote \(H_{vw}=h_{vw}e^{2i\gamma_{vw}}\)
with \(h_{vw}=|H_{vw}|\) and
\(\gamma_{vw}\in (-\frac{\pi}{2},\frac{\pi}{2}]\). Then,
\begin{equation}
  \label{phiofa}
  \phi_w = \frac{e^{i \gamma_{wv}}}{\sqrt{h_{wv}}}
  \left[a_{wv}e^{-i \pi/4 }+ a_{vw} e^{i\pi/4} \right]
  \qquad \text{and} \qquad
  \phi_v = \frac{e^{i \gamma_{vw}}}{\sqrt{h_{vw}}}
  \left[a_{wv}e^{i \pi/4 }+ a_{vw} e^{-i\pi/4}
    \right] \ .
\end{equation}
Consider a vertex \(v\) with degree \(d_v\) and the
vertices \(\{w\}\) which are connected to it. 
There are \(2 d_v\) directed edges with
\(o(e)=v\) (outgoing) or \(\tau(e)=v\) (incoming). 
The corresponding set of \(a_e\) must all satisfy
\eqref{phiofa} on all edges connected to \(v\). 
This
requirement offers \(d_v-1\) independent
homogeneous linear equations which the set of \(a_e\)
must satisfy. 
One further homogeneous linear equation
follows directly from \eqref{schrodinger_equation} by
considering the \(v\)-th row which
involves \(\phi_v\) and all the connected  \(\phi_w\). 
Thus the  set of \(2d_v\) amplitudes must
satisfy \(d_v\) equations --
which provide a
linear relation between the set of all outgoing
amplitudes \(a_{wv}\)  and the set of all incoming
amplitudes \(a_{vw}\):
\begin{eqnarray}
\label{sigv}
\mathbf{a}^{(out)}=\sigma^{(v)}(E)\mathbf{a}^{(in)} \ \ \  {\rm where} \ \ \
\sigma^{(v)}_{w'w}(E) = i
\delta_{w'w}-2
\frac{\sqrt{h_{vw'}h_{vw}}}{H_{vv}-E-i\Gamma_v}
e^{i(\gamma_{vw}+\gamma_{w'v})}\ .
\end{eqnarray}
The matrix \(\sigma^{(v)} (E)\) is a
\(d_v\times d_v\) unitary matrix for any real
\( E\). 
It depends on the matrix-elements of
the \(v\) row in \(H\). 
Combining all the vertex conditions \eqref{sigv} to a single \(D\)
dimensional matrix and observing the rule
that a directed edge \((wv) \) plays a double role --
incoming (to \(w\) from \(v\)) and outgoing
(from \(v\) to \(w\)) -- one finds that the \(D\)
dimensional amplitude vector must satisfy
\(\mathbf{a} =U(E) \mathbf{a}\). 
Hence
\(\det[\mathbb{I}-U(E)]=0\) is satisfied if and only
if \(E\) is in
the spectrum of \(H\). 
The Quantum Poincar\'e map
\(U(E)\) is a unitary matrix defined by
\begin{equation}
\label{Umatrix}
U_{v'w',vw}(E)=
\delta_{w'v}\sigma^{(v)}_{v'w}(E)\ \quad
\text{or}  \quad \ U(\lambda)=P\Sigma(E)\ ,
\end{equation}
where \(P\) is a permutation matrix and
\(\Sigma(E)\) is a block diagonal matrix
with the \(V\) diagonal blocks \(\sigma^{(v)}(E)\). 
\(U(E)\) is  unitary for any real \(E\). 
The determinant identity
\begin{equation}
  \zeta_H(E)\equiv \det[\mathbb{I}-U(E)]=\frac{2^E \det[E-H]}{
    \prod_{v=1}^V (H_{vv}-E-i
    \Gamma_v)}\ .
  \label{detid}
\end{equation}
proves that the real zeros
of \(\zeta_H(E) \) coincide with
the spectrum of \(H\) and its
poles lie in the lower half of the complex plane \cite{US2007,GnuSmi}.

\vspace*{0.5cm}
\noindent
\textbf{Step 3: Construction of the discrete classical dynamics}\\
The matrix elements of the Quantum Poincar\'e map \(U(E)\)
are the  transition amplitudes for
the discrete step. 
The absolute squares
\begin{equation}
  B_{e',e}(E)=|U_{e',e}(E)|^2
\end{equation}
are transition probabilities which define
an analogue classical dynamics in terms of a Markov process
on the underlying phase space of directed edges. 
We refer to this as the Poincar\'e-Markov map.
The classical probabilities
\(p^\mathrm{cl}_e(n)\) to be on  the directed edge \(e\) after \(n\)
time steps evolve by
\begin {equation}
  p^{cl}_{e'}(n+1) =\sum_{e \in \mathcal{D}} B_{e'e}(E)\ p^{cl}_e(n)\ .
  \label{cevol}
\end{equation} 
This `Liouvillian dynamics' is the natural classical counterpart
of the quantum mechanical description
induced by the quantum map \(U(E)\).
The  trajectories which contribute to
the transition \(e\ \rightarrow\ e'\)
in \(n\) steps are the same for
both the classical and quantum descriptions. 
However the quantum
interference obtained by summing \textit{amplitudes}  is replaced in
the classical expression  by summing transition \textit{probabilities}. 	
Comment : The matrix elements of\(B(E)\)
do not depend on the phases
of \(H_{uv}\) when \(u \neq v\).
This could be overcome partially  by constructing the stochastic
matrix from elements of \((U^2)_{uv}\) \cite{Future}.

The matrix \(B(E)\) is bi-stochastic
\(\sum_{e\in \mathcal{D}} B(E)_{ee'}= 1= \sum_{e'\in \mathcal{D}} B(E)_{ee'}\).
Bi-stochastic matrices that are obtained from a unitary matrix in an
analogue way have been called \textit{uni-stochastic} and have been
discussed in detail for quantum graphs \cite{berkolaiko,tanner,pakonski}
where an analogous quantum-classical correspondence has been used very
effectively to understand quantum chaos (see \cite{GnuSmi} and
references therein).

As such \(B(E)\) has the following  properties: 
Its spectrum, denoted by  \(\{\nu_j\}_{j=1}^D\) is
restricted to the unit disc in the complex plane,
complex eigenvalues appear in conjugate pairs. 
The uniform distribution
\(| \nu_1 \rangle =\frac{1}{{ \sqrt D}}(1,1,...,1)^T\)
is invariant, that is, it is an eigenvector with eigenvalue
\(\nu_1=1\). 
This is known as the Frobenius
eigenvalue and eigenvector and we reserve the index \(1\). 
When all eigenvalues but \(\nu_1\) are strictly inside a disc of
radius  \(1\) the evolution is mixing and the dynamics decays to the
uniform
distribution exponentially fast. 

Many properties of the system can be computed in terms of the
matrix \(B(E)\) and its spectrum. 
E.g.~ the rate of
entropy production and  the probability to return to the starting
position
after a given time. 
We focus on measures of chaoticity as expressed in
terms of the Lyapunov exponents. 

\vspace*{0.5cm}
\noindent
\textbf{Step 4: The Lyapunov Exponent}\\
In the present context a trajectory is just a sequence of
connected edges:
\begin{eqnarray}
  \xi =\left \{
  (e_t)_{t=-\infty}^{\infty}:e_t\in\mathcal{D},t\in \mathbb{Z},
  \tau(e_t)=o(e_{t+1})
  \right \}\ .\label{trajectory}
\end{eqnarray}
It can be considered as a string picked up from a collection of $D$ letters.
The propagation along the itinerary  is described by the  \textit{shift operation}
\(e_t \mapsto e_{t+1}\). 
Systems with trajectories which follow the above definitions
are called  \textit{shifts of finite type} and
are abundantly studied in ergodic theory
\cite{Gaspard, ergodictheory,Parpot,Sarig}. 

Consider a finite section of a trajectory:
\(\xi_t= (e_j)_{j=0}^{t}\) which start at a prescribed directed edge
\(e_0\). 
The probability that this trajectory will be traversed in the
stochastic evolution induced by \(B(E)\) is
\begin{equation}
	P(\xi_t)=\prod_{j=1}^{t}  B_{e_{j},e_{j-1}}\ .
\end{equation} 
The mean Lyapunov exponent 
is defined as
\begin{equation}
\langle \Lambda(E) \rangle =
-\lim_{n
t\rightarrow \infty}
\frac{1}{t}
\left\langle \log[P(\xi_t)]\right \rangle_{\xi_t}
\label {deflyap}
\end{equation}
where the average is over all trajectories of length \(t\) and all
initial \(e_0\). 

The thermodynamic formalism  provides
powerful methods to compute
the Lyapunov exponent. 
This is done by introducing an auxiliary \(D\times D\)  matrix
\begin{eqnarray}
  Q_{e,e'} (\beta)= [ B(E)_{e,e'}]^{\beta}, \ \    \beta    \ge 0\ .
\end{eqnarray} 
Denoting the eigenvalue of \(Q_{e,e'}(1+\epsilon)\) with the largest real part
by \(\mu(\epsilon)\),
one finds
\begin{eqnarray}
  \langle \Lambda(E)\rangle
  =
  -
    \left .
    \frac{
    \partial
    \log [\mu(\epsilon)]
    }{
    \partial \epsilon
    }
    \right |_{\epsilon =0}\ ,
 \qquad \text{and} \qquad
  \langle  \Lambda(E)^2\rangle
  -
  \langle \Lambda(E) \rangle ^2
  =
  \left .
    \frac{
    \partial^2
    \log [\mu(\epsilon)]
    }{
    \partial \epsilon^2
    } \right |_{\epsilon =0}\ .
\end{eqnarray}
A simple computation shows
\begin{equation}
  \langle \Lambda(E) \rangle= -\frac{1}{D}
  \sum_{e, e'\in \mathcal {D}}
  B(E)_{e,e'}\log B(E)_{e,e'}\ .
  \label{Lyap}
\end{equation}
The expression for the second moment is
quoted here for the sake of completeness
\begin{equation}
  \label{variance}
  \langle \Lambda^2(E)\rangle
  =
  \frac{1}{D}\sum_{e,e'\in \mathcal {D}} B(E)_{e,e'}( \log B(E)_{e,e'} )^2
  +2\sum_{k\ne 1}\frac{\left|\langle 1|G(E)|k\rangle\right|^2}{1-\nu_k}
  \ , 
\end{equation}
where the \(D\times D\) matrix \(G(E)\) is defined as
\(G(E)_{e'e} := B(E)_{e,e'} \log B(E)_{e,e'} \).
The detailed derivation and discussion of the
variance  is deferred to a forthcoming
publication \cite{Future}. 

More detailed information is obtained by the \textit{local} Lyapunov
exponents which measure the spread of trajectories
at a directed edge \(e\) or a vertex \(v\):
  \begin{eqnarray}	
    \langle\Lambda_{e}(E)\rangle= - \sum_{e':o(e')=\tau(e)}  B_{e' e} \log B_{e' e}
    \qquad \text{and}   \qquad
    \langle\Lambda_{v}(E)\rangle=
       \frac{1}{d_v} \sum_{e:\tau(e)=v}
      \langle\Lambda_{e}(E)\rangle,
\end{eqnarray}
where the first sum goes over the set of
directed edges  starting from the terminus of \(e\)
and the second sum is
over the directed edges
terminating at the vertex \(v\). 

When considering matrices which
correspond to
bipartite graphs
(e.g., finite trees or linear graphs as
associated with tridiagonal matrices)
the matrix \(B(E)\) is ergodic but not
mixing since \(-1\) is in the spectrum. 
To restore the stronger property of mixing,
one uses the fact that the underlying  \(U(E)\)
matrix can be decomposed to four square
blocks each of dimension \(D/2\), with vanishing two
diagonal blocks and unitary off-diagonal blocks
denoted by \(U_u (E)\)
and \(U_d (E)\). 
The spectral secular equation reads
\(\det (\mathbb{I}-U_u (E)U_d (E))=0\). 
The
unitary matrix
\(\tilde  U(E)=U_u(E)U_d(E)\)  can be used to define a
bistochastic matrix
\(\tilde B\) which has the eigenvalue \(-1\) removed. 
We will use this below for the example of the
tridiagonal G\(\beta\)E ensemble. 

The Lyapunov exponents
obey the inequalities
\begin{equation}
	\hspace {-15mm}
  0\le  \langle\Lambda_{e}(E)\rangle \le \log d_{\tau(e)}
  \qquad \text{and} \qquad
  0\le  \langle\Lambda_{v}(E)\rangle \le \log d_{v}
  \qquad \Rightarrow\qquad
  0\le  \langle\Lambda(E)\rangle \le \frac{1}{D} \sum_{v=1}^V
  d_v \log d_v\ .
\label{upper_bounds}
\end{equation}
the first two are for the local and the last for the full  Lyapunov exponents. The bounds are expressed in terms of the degrees \(d_{\tau(e)}\) and \(d_v\) of
the vertices \(\tau(e)\) and \(v\). 
The lower bounds are obtained if
there is one directed edge that follows with probability one,
while the maximum
is achieved when all connected directed edges can be reached
with the
same probability \(1/d_{\tau(e)/v}\). 

The above upper bounds hold for arbitrary bi-stochastic matrices
that obey the connectivity of the underlying graph. 
Typical values for the bi-stochastic matrix \(B(E)\) are smaller. 
If the energy \(E\) is chosen outside the spectrum of \(H\) then
\eqref{sigv} implies
\(\langle\Lambda(E)\rangle
= O\left(\frac{\log E}{E^2}\right)\)
in the limit \(E \to \pm \infty\). 
The same applies to local Lyapunov exponents. 

\section{Instructive examples}

The following examples illustrate the application of the
the local and global  Lyapunov exponents in a few cases, and in particular demonstrate their use for quantifying the underlying chaos. 

\begin{figure}[htp]
  \hspace{-1mm}
  \includegraphics[width=.65  \textwidth]{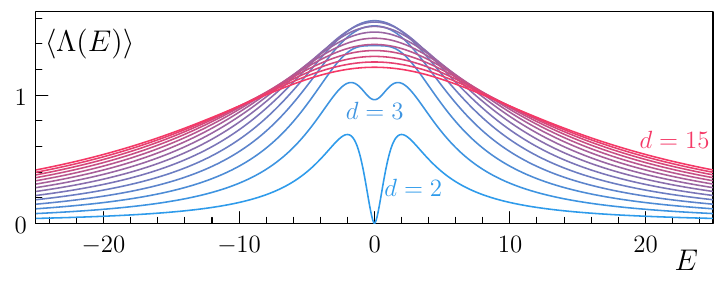}
  \caption{Mean (or, equivalently, local)
    Lyapunov exponent for \(H=A\) on a \(d\)-regular graph as a
    function of \(E\).} 
  \label{fig_regular}
\end{figure}

\vspace*{0.5cm}
\noindent
\textbf{Adjacency matrices of \(d\)-regular graphs}\\
The Lyapunov exponent takes a very simple
form if \(H\) is equal to the
adjacency matrix of a connected
\(d\)-regular graph. 
All the vertex scattering
matrices \(\sigma^{(v)}\) are then identical and the Gershgorin radius
is \(\Gamma_v=d\) for all vertices. 
At any given vertex,
the transmission and reflection probabilities are \(p_t =4/(E^2+d^2)\) and
\(p_r= 1- (d-1) p_t\). 
The mean and local Lyapunov exponents
are identical -
\(
\langle \Lambda(E) \rangle =
-p_r\log p_r
-(d-1)p_t\log p_t
\) (see Fig.~\ref{fig_regular}). 
For \(d\le 4\) the maximal Lyapunov exponent
is attained at
\(E= \pm \sqrt{d(4-d)}\) and given by the upper bound
\(\log d\) derived before in
\eqref{upper_bounds}. 
Otherwise the  maximal Lyapunov exponent occurs at \(E=0\) where
\begin{equation}
  \langle \Lambda(0)\rangle
  =\frac{ 2(d-2)^2}{ d^2}\log
  \left(\frac{ d}{ d-2}\right) +
  \frac{ 8(d-1)}{ d^2}
  \log\left(\frac{ d}{ 2} \right)\ .
\end{equation} 
For \(d>4\) this is strictly smaller than the upper bound
\(\log d\). 
For  \(d\to \infty\) one has
\(\langle \Lambda(0)\rangle \sim\frac{8 \log(d)}{d} \ll \log d\).
These maximal values may be used as benchmarks for
local and global Lyapunov exponents in general. One should then
 replace
\(d\) by the mean degree of the graph
and any degree of non-uniformity
decreases the Lyapunov exponent further.
It also shows that
comparing actual values of Lyapunov exponents on different
graph structures needs to be performed with care. 
It is also worth mentioning that the independence of the local and
the mean Lyapunov exponent on $V$ or the detailed graph
connectivity does not persist to the variances \cite{Future}.

\begin{figure}[htp]
  \hspace{-1mm}
  \includegraphics[width=.65  \textwidth]{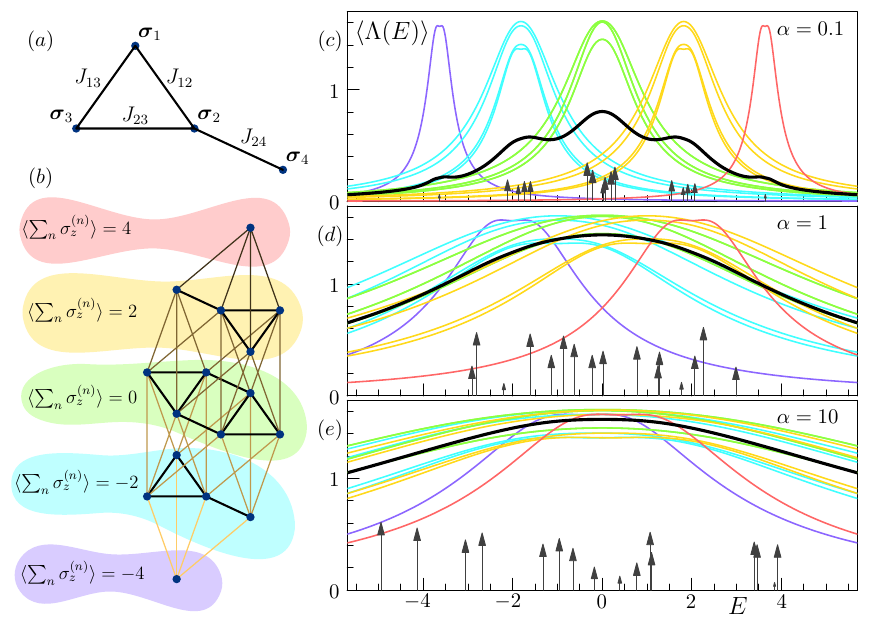}
  \caption{(a) Spin graph with \(V_{\mathrm{spin}}=4\)
    vertices.\\ 
    (b) Corresponding
    graph of the Hamiltonian where the 16 vertices correspond to
    spin configurations.\\ 
    (c-e) Mean Lyapunov exponent (black), local
    Lyapunov exponents (coloured lines,
    colours correspond to the ones used in (b)). 
    The spectrum is located at the black arrows where the height
    corresponds to the participation ratio divided by \(V=16\). 
    We show results for \(J_{12}=\frac{1}{3}\), \(J_{13}=\frac{\sqrt{5}}{3}\),
    \(J_{23}=\frac{\sqrt{11}}{3}\),
    and  \(J_{24}=\frac{1}{\sqrt{3}}\). 
    }
  \label{fig_spingraph}
\end{figure}

\vspace*{0.5cm}
\noindent
\textbf{A spin-graph Hamiltonian}\\
This example
shows how the combination of the local and global Lyapunov exponents
provides a tool for
analysing the dynamics under study. 
Consider \(V_{{\rm spin}}\) spins \(\boldsymbol{\sigma}^{(v)}\)
attached to vertices on a graph
that interact pairwise according to the connectivity of the graph with real
coupling strengths \(J_{vw}=J_{wv}\). 
The Hilbert space of dimension
\(V=2^{V_{\mathrm{spin}}}\) is spanned by product
states with \(\sigma_z^{(v)}\) eigenvalues
equal to \(\pm 1\). 
All spins are subject to a homogeneous magnetic
field. 
For definiteness we choose a Hamiltonian
\begin{equation}
  \begin{split}
    &H= {\frac{1}{1+\alpha}}H_0 +
    {\frac{\alpha}{1+\alpha}} H_I , \qquad
    \text{where}\\
    &H_0=
    \sum_{v} \sigma_z^{(v)}\ ,
    \qquad \text{and} \qquad
    H_I=
    \sum_{v<w} J_{vw}{
      (\sigma_x^{(v)}\sigma_x^{(w)}
      +\sigma_y^{(v)}\sigma_y^{(w)}  +\sigma_x^{(v)}\sigma_z^{(w)}+
      \sigma_z^{(v)}\sigma_x^{(w)} )}
  \end{split}
\end{equation}
and
\(\alpha>0\) controls the relative strength. 
In Fig.~\ref{fig_spingraph} we show the mean and
local Lyapunov
exponents for a particular choice of the spin graph and some values of
the interaction strength \(\alpha\). 
The figure shows that the mean Lyapunov exponent remains much smaller
than the maximal
local one for a weak
coupling where eigenstates remain mainly within a subspace of
constant \(\langle \sum_v \sigma^{(v)}\rangle \) while
stronger couplings lead to more uniform distributions. 

\begin{figure}[htp]
 	\includegraphics[width=.65  \textwidth]{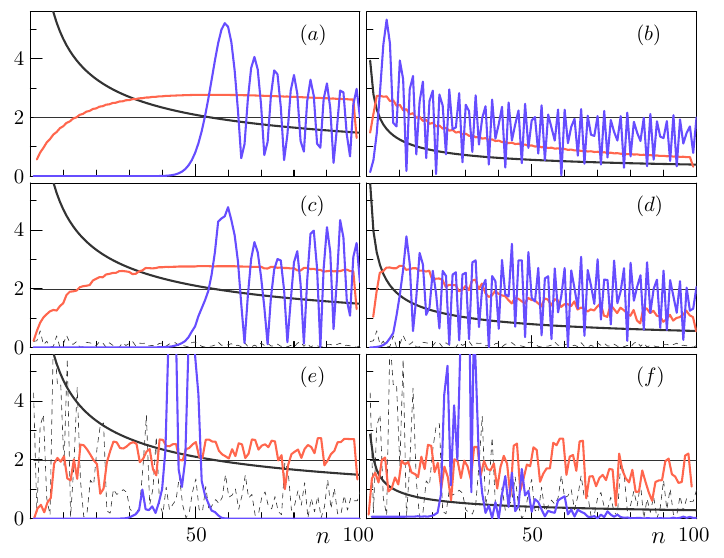}
 	\caption{
          Squared wavefunctions (Blue, arbitrary scale),
          local Lyapunov exponents (Red, arbitrary scale),
          and effective potential \(W_{\mathrm{eff}}(n)\)
          (Black). 
          (a-b) Mean field behaviour for a
          high (a) and low (b)
          value of the energy \(E\) within the spectrum. 
          (c-d)
          Single realisation  at \(\beta=10\)
          (the dashed line gives the random potential). 
          (e-f) The same for \(\beta =0.1\). } 
 	\label{figGbetaE}
 \end{figure}

\vspace*{0.5cm}
\noindent
\textbf{The G\(\beta\)E ensemble of tridiagonal matrices}\\
The  G\(\beta\)E ensemble
\cite{Edelman} of \(V \times V\)
tridiagonal matrices offers a simple case for  studying
the local Lyapunov exponent when the Hamiltonian
consists of independently distributed random entries. 
The diagonal \(H_{nn}=a_n\)
are distributed normally with zero mean and variance \(1\). 
The off-diagonal elements \(H_{n-1,n}= H_{n, n-1}=b_n\) are
distributed with
\(p(b_n) = \frac{2}{\Gamma(\frac{\beta n}{2})}b_n^{\beta n-1}\exp[-b_n^2]\). 
\(\beta>0\)
is a parameter which characterise the ensemble. 
The spectral statistics coincides with the counterpart
Wigner-Dyson ensembles for \(\beta =1,2,4\).
For large \(n\) one finds to leading order
\begin{eqnarray}
  \langle b_n\rangle \sim \sqrt{\frac{\beta n} {2}}(1-\frac{1}{4\beta n}) \qquad \text{and}  \qquad
  \langle (b_n - \langle b_n\rangle)^2\rangle \sim\frac{1}{4}\ . 
\end {eqnarray}
While the mean of \(b_n\) is growing as \(\sqrt{\beta n}\),
its variance tends to a constant. 
Hence, by increasing \(\beta\), the mean value of the off-diagonal entries
become increasingly dominant, and the effect of fluctuations diminish. 
This leads to the mean field  discrete Schr\"odinger equation
\begin{equation}
  \sqrt{\frac{\beta}{2}}\left((n-1)^{\frac{1}{2}}\phi_{n-1}
    + n ^{\frac{1}{2}}\phi_{n+1}\right )=E \phi_n
  \label{rec}
\end{equation}
with \(\phi_0=\phi_{N+1}=0\). 
For \(g(n)=n^{\frac{1}{2}}\phi_n\) the ODE analogue of \eqref{rec} reads
\begin{eqnarray}
	-\frac{d^2 g(n)}{d n^2} +\sqrt{\frac{2}{\beta n}}E g(n)= 2g(n) \ .
\end{eqnarray} 
It describes a particle with ``energy'' \(2\) subject to a
potential \(W_{\mathrm{eff}}=\sqrt{\frac{2}{\beta n}}E\),
with classical turning point  at  \(n_t = \frac{E^2}{2\beta}\). 
Fig.~\ref{figGbetaE}(a-b) show the absolute square values of
the eigenfunctions and the Lyapunov exponents
for \eqref{rec} for two eigenvalues \(E\) which belong to the higher  (left)
and lower (right) parts  of the spectrum. 
The main feature is the appearance of
domains on the \(n\) axis where the eigenvector amplitudes are small.
The domains  starts
at \(n=1\) and extend up to the classical turning point $n_t$. This is the classically forbidden domain.
The local Lyapunov exponents
follow this behaviour,
indicating that the phenomenon is captured by the underlying classical
dynamics. 
This feature was studying in detail in \cite{Sodin}.

Fig.~\ref{figGbetaE}(c-f) show results for individual realisations of the ensemble
for comparison. 
Fig~\ref{figGbetaE} (c-d) are computed for \(\beta =10\). 
The wave functions and
the local Lyapunov exponents
are rather similar to their counterpart shown in Fig.~\ref{figGbetaE}(a-b),
albeit more noisy. 
Fig~\ref{figGbetaE} (e-f) are  computed for \(\beta =0.1\). 
It displays a radically different behaviour because for
low values of \(\beta\) the random potential dominates
leading to localisation of
the wave functions. 
Note that the dependence on \(\beta\) is through its square root,
so the actual effect of changing \(\beta\) is only a factor \(10\). 
Averaging over many realisations the Lyapunov exponents
approach the results obtained
for the mean potentials. 
Thus, as long as the  classical effects dominates in determining the dynamics,
the local Lyapunov exponents follow the behaviour of the eigenvectors. It fails when interference  dominates and in particular when localisation is not due to the presence of classically forbidden domains.  


\section{Discussion and Outlook}
Two issues will be briefly addressed:
\textbf{(a)}
the connection between the present approach and the LR bound and the OTOC;
\textbf{(b)}
a short list of subjects that should be further addressed and open problems.

\noindent
\textbf{(a)} Both the LR and the OTOC methods use the time dependent commutator
\([A(t),B]\), where
\(A(t)\equiv  e^{-\frac{i}{\hbar}Ht} Ae^{ \frac{i}{\hbar}Ht}\)
and \(B\) are Hermitian operators selected for the
purpose. They show that the norm of the commutator increases (at most)
exponentially
under some conditions and the exponential growth rate is the
indicator of the rate of scrambling induced by \(H\).
In the discrete time setting used here we replace the continuous
time propagator by \(U^t(E)\) and for the sake of clarity we
use  \(A =|a\rangle\langle a|\) and \(
B=|b\rangle\langle b|\) which project on two directed edges a distance
\(L\) apart. Choosing the Frobenius operator norm, and
abbreviating  \(U(E)\) by \(U\) one obtains after some simple
computation 
\begin{equation}
\label{otoc}
\frac{1}{D} { \rm tr} 	\left ( [U^t A U^{t^{\dagger}},B][U^t A U^{t^{\dagger}},B]^{\dagger} 
\right ) =\frac{2}{D} \left ( |(U^t)_{ba}|^2 - |(U^t)_{ba}|^4  
\right ) .
\end{equation}
Denoting by \(\xi _t\) the trajectories \eqref{trajectory}
connecting the edges \(a\) and \(b\) by traversing \(t\) connected edges,
and the set of these trajectories by \(\Xi_t\)
(the number of trajectories   is denoted by \(|\Xi_t|\) )
we get for \(t\ge L\)
\begin{equation}
0\le |U^t_{ba }|^2=\left |\sum_{\xi\in  \Xi_t} \prod_{i=1}^{t+1}U_{e_i e_{i-1}}\right |^2  
=\sum_{\xi\in  \Xi_t} \prod_{i=1}^{t+1}B_{e_i e_{i-1}} +\sum_{\xi\ne \xi'\in  \Xi_t} \prod_{i=1}^{t+1}U_{e_i e_{i-1}}\prod_{i'=1}^{t+1}U^{\dagger}_{\ e_{i'} e_{i'-1}} \le 1 \ .
\end{equation}
For \(L<t<D\)
the last sum consists of highly oscillatory terms and
therefore can be neglected under appropriate averaging, e.g. with respect to
a time window.
(It cannot be neglected for \(t\) close to \(D\) or larger because of the
occurrence of increasingly larger number of trajectories
which visit the same edges in different orders and therefore
have exactly the same phases.
This double sum then keeps the expression bounded by \(1\).)
The term \(|(U^t)_{ba}|^4\) in \eqref{otoc} can also safely be neglected. 
Using the fact that the geometric mean is always smaller than
the arithmetic mean, one gets 
\begin{equation}
  |\log[ |U^t_{ba }|^2]| \le t \left\{  \frac{1}{|\Xi_t|}
    \sum_{\xi\in  \Xi_t}   \frac{1}{t}
    \sum_{i=1}^{t+1}|\log [ B_{e_i e_{i-1}} ]|\right \} -\log[|\Xi_t|]
	\label{coef}
\end{equation}

The curly sum appearing in \eqref{coef} is the analogue of the
equation defining the the mean Lyapunov exponent \eqref{Lyap},
the only difference being that the averaging is different.
In the present case all trajectories are treated with equal
probability where as in \eqref{Lyap} the probabilities of the
trajectories are used as weights. 

The main advantages of the present formalism is that it allows the
computation of the variance of the Lyapunov exponent, and that
the trajectories are assigned with an energy \(E\).
Practically, the computation of the mean Lyaponov exponent does not
require the spectrum or any matrix inversion.

\noindent
\textbf{(b)} The open problem which are left for forth coming
paper \cite{Future} are to test the method on systems which are closer
to the field where OTOC is applied. It also crucial to investigate
to what extent the Lyapunov exponent depends on the spectral
parameters of the stochastic matrix, and in particular on the
gap between the leading value 1 and the rest. Several of the
approximation which were made in the first paragraph above should be
better studied, and some scale by which one could compare the
Lyapunov exponents of different systems should be established. 


\begin{acknowledgments}
  We are indebted to Omri Sarig for clarifying
  several issues in ergodic theory and to
  Sasha Odin for discussing some of his new results on the G\(\beta\)E
  model before publication,
  to Hans Weidenm\"uller who recommended corrections to an early
  draft of the manuscript, to Holger Schanz and Gregor Tanner for their comments on a later draft,
  to  Micha Berkooz for discussing the relation to the OTOC approach,
  and to  Tomas  Maci\k{a}\.{z}ek  and to Yiyang Jia for sharing with
  us their many spins models.
\end{acknowledgments}


\end{document}